# High-temperature phonons in h-BN: momentum-resolved vibrational spectroscopy and theory

Andrew O'Hara[1]†‡, Benjamin Plotkin-Swing[2]*†, Niklas Dellby[2], Juan Carlos Idrobo[3], Ondrej L. Krivanek,[2,4] Tracy C. Lovejoy[2], and Sokrates T. Pantelides[1,5]*

[1]*Department of Physics and Astronomy, Vanderbilt University, Nashville, TN 37235, USA*
[2]*Nion R&D, Kirkland, WA 98034, USA*
[3]*Department of Materials Science and Engineering, University of Washington, WA 98195, USA*
[4]*Department of Physics, Arizona State University, Tempe, AZ 85287, USA*
[5]*Department of Electrical and Computer Engineering, Vanderbilt University, Nashville, TN 37235, USA*

**Abstract:**

Vibrations in materials and nanostructures at sufficiently high temperatures result in anharmonic atomic displacements, which leads to new phenomena such as thermal expansion and multiphonon scattering processes, with a profound impact on temperature-dependent material properties including thermal conductivity, phonon lifetimes, nonradiative electronic transitions, and phase transitions. Nanoscale momentum-resolved vibrational spectroscopy, which has recently become possible on monochromated scanning-transmission-electron microscopes, is a unique method to probe the underpinnings of these phenomena. Here we report momentum-resolved vibrational spectroscopy in hexagonal boron nitride at temperatures of 300, 800, and 1300 K across three Brillouin zones (BZs) that reveals temperature-dependent phonon energy shifts and demonstrates the presence of strong Umklapp processes. Density-functional-theory calculations of temperature-dependent phonon self-energies reproduce the observed energy shifts and identify the contributing mechanisms.

*Corresponding authors: pantelides@vanderbilt.edu, plotkin-swing@nion.com

†These authors contributed

‡Present address: Department of Physics, Western Michigan University, Kalamazoo, MI 49008, USA

**Main Text:**

Phonon-excitation spectra and phonon dispersions are traditionally measured using infrared, Raman, X-ray, and neutron scattering, which have limited spatial resolution and require relatively large sample volumes. Over the last twenty years, aberration-corrected scanning-transmission-electron microscopy (STEM) for atomic-resolution imaging and electron energy-loss spectroscopy (EELS) have proved to be powerful tools for studying the structure, chemical composition, and electron excitations in solids using nanoscale samples. More recently, a new generation of monochromators and spectrometers has improved the energy resolution of EELS from the intrinsic ~250-meV energy dispersion of the electron gun to only a few meV, opening the door for STEM-based vibrational spectroscopy (*1*). STEM instruments can be operated in two complementary imaging and spectroscopy modes, either with high spatial or high momentum resolution. High spatial resolution has led to mapping phonons in materials and nanostructures at surfaces (*2*, *3*), interfaces (*4*–*6*), and grain boundaries (*7*), atom-by-atom vibrational spectroscopy at impurities in graphene (*8*, *9*), and even the detection of isotopes at the nanoscale (*10*–*12*). Trading high spatial resolution for momentum resolution permits the mapping of phonon dispersions across several Brillouin zones (BZ), including low-energy acoustic and optical modes (*13*–*16*). Combining these experimental probes with density-functional-theory (DFT) calculations and simulations has provided mutual validation of the pertinent analysis of vibrational properties.

Investigation of phonons in materials at high temperatures using STEM/EELS remains largely unexplored. So far, it has only been demonstrated that, by using the principle of detailed balance, STEM can act as a thermometer, measuring the temperature of materials via EELS with nanometer spatial resolution (*17*, *18*). The high-temperature regime, however, enables access to a wide range of phenomena that involve phonons. At high thermal excitations, which in many materials can occur near room temperature (*19*, *20*), larger atomic displacements break outside the harmonic limit and induce significant anharmonic-phonon effects. Typical consequences are thermal expansion (*20*, *21*) and multiphonon scattering processes (*20*, *22*), which lead to significant renormalization of phonon energies with profound impact on temperature-dependent material properties including thermal conductivity, phonon lifetimes, and phase transitions. In complex materials and nanostructures these phenomena can vary on nanoscale dimensions.

In this article, we use momentum-resolved EELS in monochromated aberration-corrected STEM to investigate the temperature dependence of the phonon dispersion and spectra of hexagonal boron nitride (h-BN), a layered van der Waals material, across three BZs at 300, 800, and 1300 K. Sufficient narrowing of the zero-loss peak by the monochromator allows all acoustic and optical in-plane phonon modes and their dispersions to be observed. Temperature-dependent anharmonic phonon softening, while small, is observable. Density-functional-theory calculations of phonon energy shifts, including both thermal expansion and many-body phonon-phonon effects, provide mutual validation of both the direction and magnitudes of the calculated and experimentally observed anharmonic changes. Furthermore, the calculations provide insights into the microscopic origins of the anharmonic shifts, namely cubic *vs*. quartic anharmonicity and three-phonon *vs*. four-phonon interactions. In addition to phonon energy shifts, significant evolution in the EELS relative mode intensities is observed for momentum transfers outside the first BZ and with increasing temperature. Though, as expected, energy shifts at equivalent momenta within and outside the first BZ are identical, the corresponding relative EELS phonon intensities differ significantly, signaling the presence of strong Umklapp processes. Currently available theory based on the Stokes cross-section, which does not include Umklapp contributions, gives a satisfactory description of the relative EELS phonon intensities only at room temperature and momentum transfers within the first BZ, indirectly confirming the presence of strong Umklapp contributions at large momentum transfers.

Hexagonal BN has large out-of-plane thermal expansion but small, negative in-plane expansion (*23–25*), reflecting highly anisotropic phonons in the in- and out-of-plane directions. Anharmonic effects are known to be significant, especially at higher temperatures (*26*). Due to the approximately orthogonal beam-sample geometry, at finite momenta away from the Γ-point, only modes derived from the in-plane transverse acoustic (TA), longitudinal acoustic (LA), transverse optical (TO), and longitudinal optical (LO) modes are observed [the so-called ZA- and ZO-based modes can be detected if the sample is tilted relative to the beam (*15*)]. Overall, h-BN is a good test case to explore the new EELS capabilities and to jointly probe with theory phonon properties at elevated temperatures.

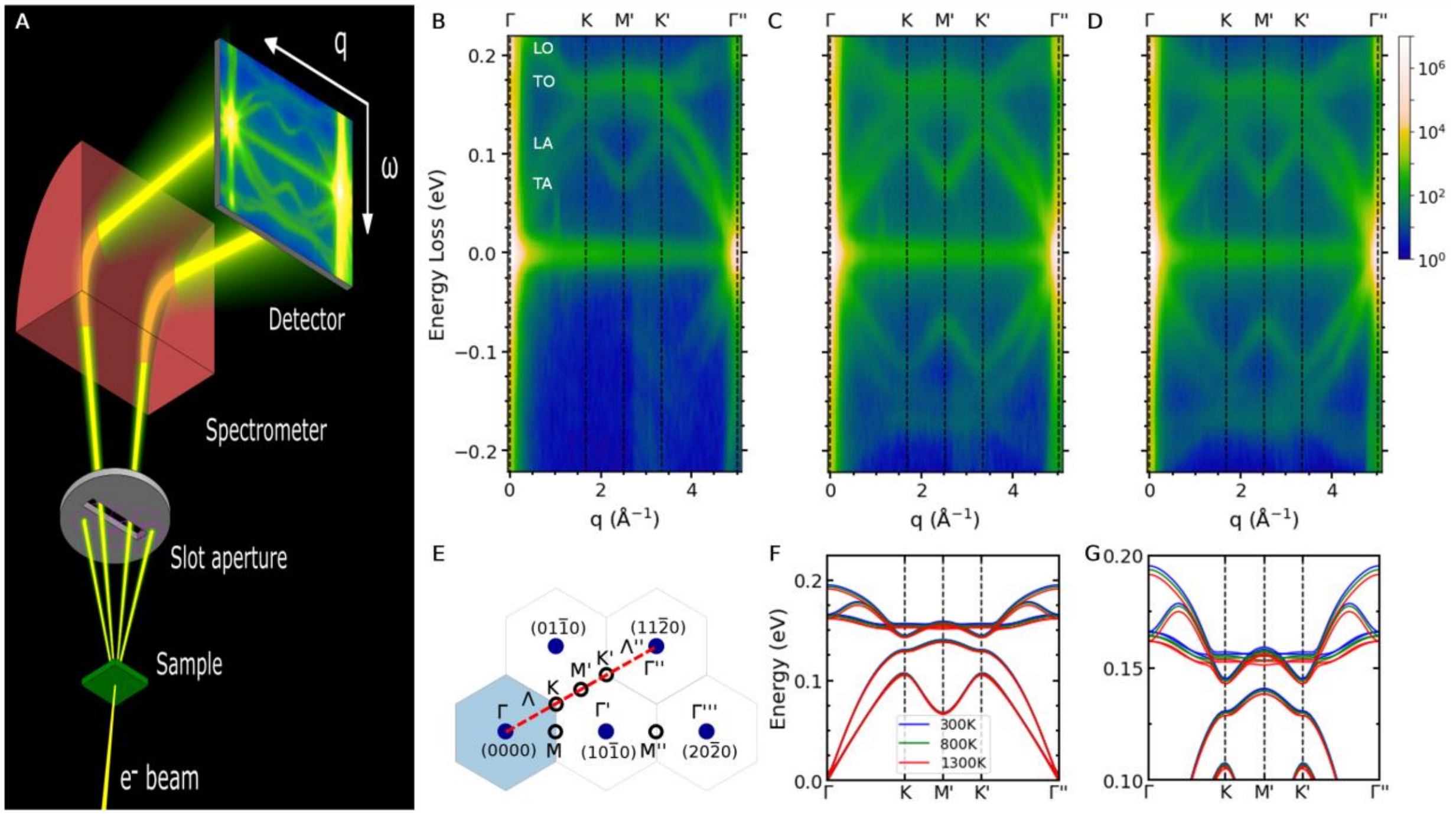

**Figure 1. (A)** Principle of the experimental set-up. About 6-nm wide incoming electron beam is scattered by a thin sample and a line in the resultant diffraction pattern is selected by a slot aperture. The selected beams pass through the spectrometer, which disperses them in the energy direction, and a momentum-energy (q-ω) pattern that analyzes a line section through one or more Brillouin zones is projected onto the detector. (**B-D**) STEM-EELS-derived Phonon dispersions derived from STEM/EELS energy loss (upper halves) and energy gain (lower halves) at 300 K (**B**)**,** 800 K (**C**), and 1300 K (**D**) along the high-symmetry momentum-transfer path connecting the first and third BZs, from Γ to Γ″, shown in the diagram in (**E**). (**F**) Calculated phonon dispersion for each temperature along the same path. **(F)** Zoom-in of the energy axis in (**G)** to better illustrate the optical-mode shifts.

Figure 1A shows a schematic diagram of the apparatus used for the present experiments. The 60 keV electron beam incident on the sample is formed by aberration-corrected optics (*27*) and a high-stability monochromator (*28*, *29*). The q-ω pattern is recorded in parallel by a hybrid pixel direct detector (*30*). It is typically 5-10 meV wide in energy and the spatial and angular resolutions are diffraction-limited to about 6 nm size at the sample and 1 mrad angular resolution. For simplicity, no lenses and only a few diffracted beams are shown in the schematic. The q-ω pattern is shown as it would emerge from the prism: upside-down relative to how it is normally presented. In reality, the total apparatus uses twelve round electromagnetic lenses, four energy-dispersing prisms, more than forty strong quadrupoles, and many sextupoles and octupoles. This configuration allows it to form a monochromatic and aberration-corrected probe of adjustable angular width on the sample, to rotate the diffraction pattern on the slot aperture so that the desired direction in the Brillouin zone can be analyzed, and to vary the dispersion in angle and energy. Typical acquisition times per pattern are 10-20 minutes – a major improvement over the many hours needed to acquire q-ω patterns in a serial manner (*15*). Further details regarding the current experimental preparations are provided in the Supplementary Materials.

In mapping the phonon dispersion as a function of momentum (Fig. 1B-1D), we follow a path along the $(11\bar{2}0)$ direction from the Γ point in the first BZ to the Γ″ point in the third BZ passing through the K, M′, and K′ high-symmetry points along the edge of the second BZ as illustrated in Fig. 1E. A piecewise affine transformation was applied to the collected momentum range to correct for a systematic scaling effect (see Methods section of the Supplementary Materials and Fig. S1). Comparing the dispersions at the three temperatures, we see that all expected mode types (TA, LA, TO, LO) are visible across the momentum-transfer range. The dispersions above ~100 meV at 300 K (Fig. 1B) are in agreement with the reported EELS-derived dispersions in the Extended Data Fig. 3 of Ref. (*15*) in this energy range and with the same momentum cut.

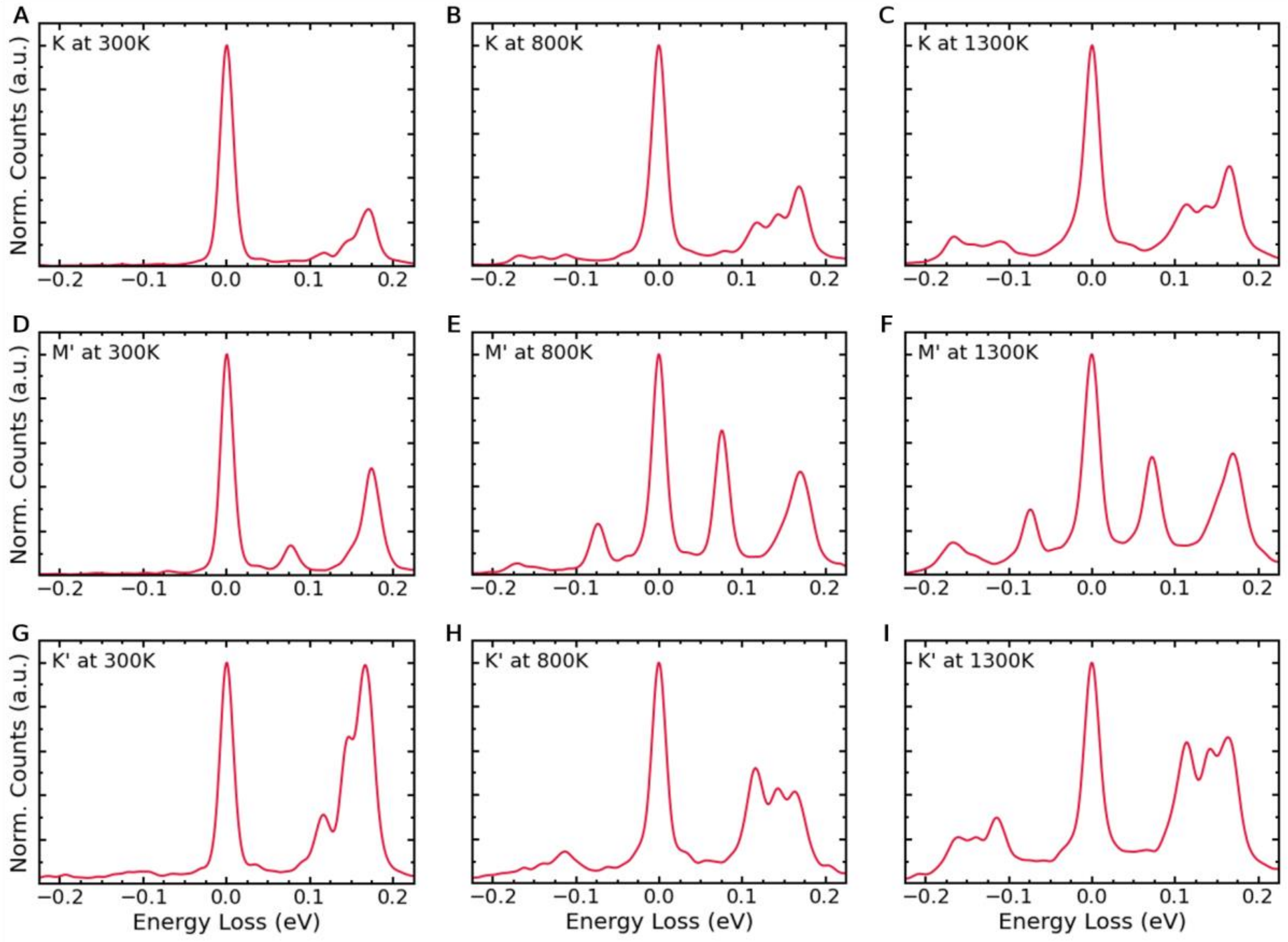

**Figure 2.** Experimental STEM-EEL spectra in the full energy-gain/loss at high-symmetry BZ points K, M′, and K′ at 300, 800, and 1300 K, as indicated in each panel. Negative energy values correspond to energy gain. The relative intensities at different temperatures are not meaningful because the three data collections are distinct experimental events so that teach spectrum is normalized at the zero-loss peak.

In Fig. 1F and 1G, we show the calculated dispersion of in-plane phonon modes at each of the considered temperatures. Thermal-expansion and anharmonic effects were included at each temperature. Anharmonicities were included via the self-consistent phonon theory (SCP) (*31*, *32*), a many-body treatment beyond standard perturbation theory, with the renormalized phonon energies obtained self-consistently through the phonon self-energy. Cubic and quartic anharmonicities with three- and four-phonon interactions were included. For further details, see Methods in the Supplementary Material. Figure 1G shows the same calculated modes, zooming on the phonon modes above 100 meV where anharmonic shifts are more pronounced.

In order to analyze the evolution of EELS data, we focus on spectra at specific momenta at high-symmetry points. In Fig. 2, we show individual spectra at all three temperatures for the BZ points K (Fig. 2A-2C), M′ (Fig. 2D-2F), and K′ (Fig. 2G-2H), showing both the energy gain (negative values on the EELS energy axis) and loss. The spectral intensity increases at the energy-gain region for each momentum transfer at higher temperatures, as more high-energy modes are thermally excited and can transfer energy to the electron beam. At 300 K, there are hardly any thermally excited modes that can contribute to the energy gain. However, at 800 K, there are clear gain contributions by all phonon modes. Furthermore, there is a clear evolution of the relative intensities of each phonon-mode contribution to the loss spectra across temperatures that we will be discussed later. In principle, the gain/loss ratio for a given mode at a given momentum can also act as a nanoscale temperature probe for the sample (*17, 18*) which we discuss in the Supplemental Materials.

Using the EEL spectra from Fig. 2, neglecting the gain components, we performed Lorenztian deconvolutions into individual mode contributions for the three different momenta and the three temperatures as shown in Fig. S3. A constrained simultaneous deconvolution was performed for the K and K′ points, where, by symmetry, the phonon energies are the same. Due to the proximity of the TO- and LO-derived energies at the M′ point, only a single averaged phonon energy can be extracted. The extracted phonon-mode energies are summarized in Table S1 (the full fitting parameters are given in Table S2). Because there is an overall systematic shift in calculated and EELS-derived phonon energies, we compare the shifts at 800 and 1300 K relative to values at 300 K. The shifts from both experiment and theory are listed in Table 1. We note that there are a total of eight in-plane phonon modes or four nearly doubly degenerate modes at each $q$ point because there are two formula units per unit cell (known as a Davydov splitting). Here, the Davydov splittings are quite small (< 1 meV) and, therefore, we averaged the calculated values as the splittings are not detectable by EELS.

In general, the phonon energy shifts shown in Table 1 indicate satisfactory agreement between experiment and theory for both the K/K′ and M′ points. For the K/K′ comparison, the notable exception is the disagreement for the TA mode. A potential reason for this disagreement is that the TA modes at these momenta are both closer to the Γ point and lower in energy (i.e., closer to the ZPL), both of which lead to an increased susceptibility to noise. For the M′ point, a notable disagreement occurs for the LA-derived mode shifts, likely arising from the fact that the mode appears as a very weak shoulder on the much-higher-intensity optical mode. For the TA mode at the M′ point, the overall separation of this mode from the others and its relative intensity indicates that it may be the theoretical calculation that underestimates the shift. Overall, however, these comparisons support the ability to extract anharmonic phonon shifts caused by high-temperature effects from EELS and that theory generally corroborates this sensitivity.

| Momentum | Temperature | TA | | LA | | TO | | LO | |
|---|---|---|---|---|---|---|---|---|---|
| | | Exp. | Th. | Exp. | Th. | Exp. | Th. | Exp. | Th. |
| K/K′ | 800 K | +1 | -0.4 | -3 | -0.8 | -1 | -0.9 | -2 | -1.6 |
| | 1300 K | -2 | -1.6 | -4 | -1.9 | -5 | -2.2 | -4 | -3.3 |
| M′ | 800 K | -2 | -0.2 | +2 | -1.0 | -3 | -1.5 | | |
| | 1300 K | -5 | -0.4 | +6 | -2.4 | -3 | -3.4 | | |

**Table 1.** Phonon energy shifts, in meV, of each of the TA-, LA-, TO-, and LO-derived phonon modes at K/K′ and M′ at 800 and 1300 K relative to 300 K. Due to the closeness in energies, only a combined TO/LO value is given for the M′ point.

Next, we examine the various contributions to the anharmonicity that controls the energy shifts as the temperature is increased from 300 to 800 and 1300 K. In our comparison between the EELS-derived and calculated phonon shifts, full treatment of the anharmonicity up to and including four-phonon cubic and quartic effects were taken into account. However, by breaking down the origin of contributions, we are able to gain additional insights into the basic material properties.

We concentrate on the longitudinal acoustic (LA) and longitudinal optical (LO) modes at the K/K′ points. We then dissect the shifts into the following components, as outlined in Table 2 (see Supplementary Material for additional details): thermal lattice expansion (LTE), as well as quartic and cubic anharmonicities, differentiated by whether only three-phonon or both three- and four-phonon scattering are considered. In every scenario, LTE results in a positive shift, counteracting the observed, ultimately, negative values observed.

| Mode | T | LTE | Quartic (3 ph) | Quartic + Cubic (3 ph) | Quartic + Cubic (3 and 4 ph) |
|---|---|---|---|---|---|
| LA, K/K′ | 800 K | +0.4 | +0.0 | -0.0 | -0.8 |
| | 1300 K | +0.5 | -1.0 | -1.3 | -1.9 |
| LO, K/K′ | 800 K | +0.6 | -1.4 | -1.7 | -1.6 |
| | 1300 K | +0.7 | -2.3 | -3.4 | -3.3 |

**Table 2.** Distinct contributions to temperature-dependent phonon energy shifts for two modes. Each column after LTE shows the new total shift. See the text for explanation of the column headings.

For the LA mode, three-phonon quartic anharmonicities lead to a negative shift at 800 K that cancels the LTE contribution and a significant shift at 1300 K that leads to an overall negative shift of the mode. Three-phonon cubic anharmonicities are negligible at 800 K and play a minor role at 1300 K. However, inclusion of four-phonon scattering causes relatively significant shifts at both temperatures.

For the LO mode, three-phonon quartic anharmonicities are much stronger than for the LA mode at both temperatures. Additionally, the cubic anharmonicities are important at 800 K unlike for the LA mode, and are much more significant at 1300 K than for the LA mode. Conversely, for the LO mode, four-phonon scattering is unimportant.

We now examine the relative intensities of the EEL line spectra at different momentum transfers and temperatures and discuss the possible origins of their evolution. We first focus on the line energy-loss spectra at 300 K at momentum transfers corresponding to the high-symmetry BZ points K, M′, and K′. We plot these spectra in Fig. 3 without deconvolutions (which are shown in Fig. S3). We added two more momentum points, Λ and Λ″, by which we mean the mid-points of the Γ-K and K′-Γ″ lines (Λ lines) shown in the diagram of Fig. 1E. In Fig. 3, we also include a column of theoretical curves at 300 K that are based on the Stokes double-differential cross section for the phonon EELS intensity. This theory was first developed twenty years ago in the case of X-ray scattering from lattice vibrations (*33*, *34*), treating the crystal as a collection of atoms that vibrate carrying a mode-independent effective charge. The theory was recently adapted to an electron beam scattering from lattice vibrations (*14*, *15*):

$$\frac{d^2\sigma}{d\Omega d\omega}(\boldsymbol{q},\omega) = \frac{4}{a_0^2}\frac{\hbar}{q^2}\sum_\nu \frac{1+n_{\boldsymbol{q}\nu}}{\omega_{\boldsymbol{q}\nu}}\left|\sum_I \frac{1}{\sqrt{M_I}} e^{-W_I(\boldsymbol{q})}\boldsymbol{Z}_I(\boldsymbol{q})\cdot\boldsymbol{\epsilon}^I_{\boldsymbol{q},\nu} e^{-i\boldsymbol{q}\cdot\boldsymbol{\tau}_I}\right|^2 \delta(\omega-\omega_{\boldsymbol{q}\nu}). \tag{1}$$

Here, $\boldsymbol{q}$ is the momentum transfer, the $\omega$'s are phonon energies, $\nu$ labels the phonon branches, $I$ labels atoms in the unit cell, $n_{\boldsymbol{q}\nu}$ is the Bose-Einstein mode occupation factor, $M_I$ and $\boldsymbol{\tau}_I$ are atomic masses and positions, respectively, $e^{-W_I(\boldsymbol{q})}$ is the Debye-Waller factor, and $\boldsymbol{\epsilon}^{I}_{\boldsymbol{q},\nu}$ are the phonon-mode atomic polarization vectors. $\boldsymbol{Z}_I(\boldsymbol{q})$ are mode-independent effective charges for lattice vibrations of momentum $\boldsymbol{q}$, approximated by a rigid-ion model using atomic form factors (*14*) or through density-functional perturbation theory (*15*).

In the right-hand column of Fig. 3, we show EEL spectra calculated at 300 K using Eq. (1) as in Ref. (*15*) for each of five BZ momentum points (see Methods in Supplementary Material). In order to identify the role of the Fermi-golden-rule matrix element in Eq. (1), we set the sum inside the modulus squared to a constant, which results in the blue curves in the five panels (the prefactor terms are still included). The latter are essentially single-momentum phonon densities of states that include the Bose-Einstein occupation term and the inverse energy dependence. They clearly demarcate the theoretical peak positions, which remain unchanged in the red curves, which include the matrix elements as in Eq. (1).

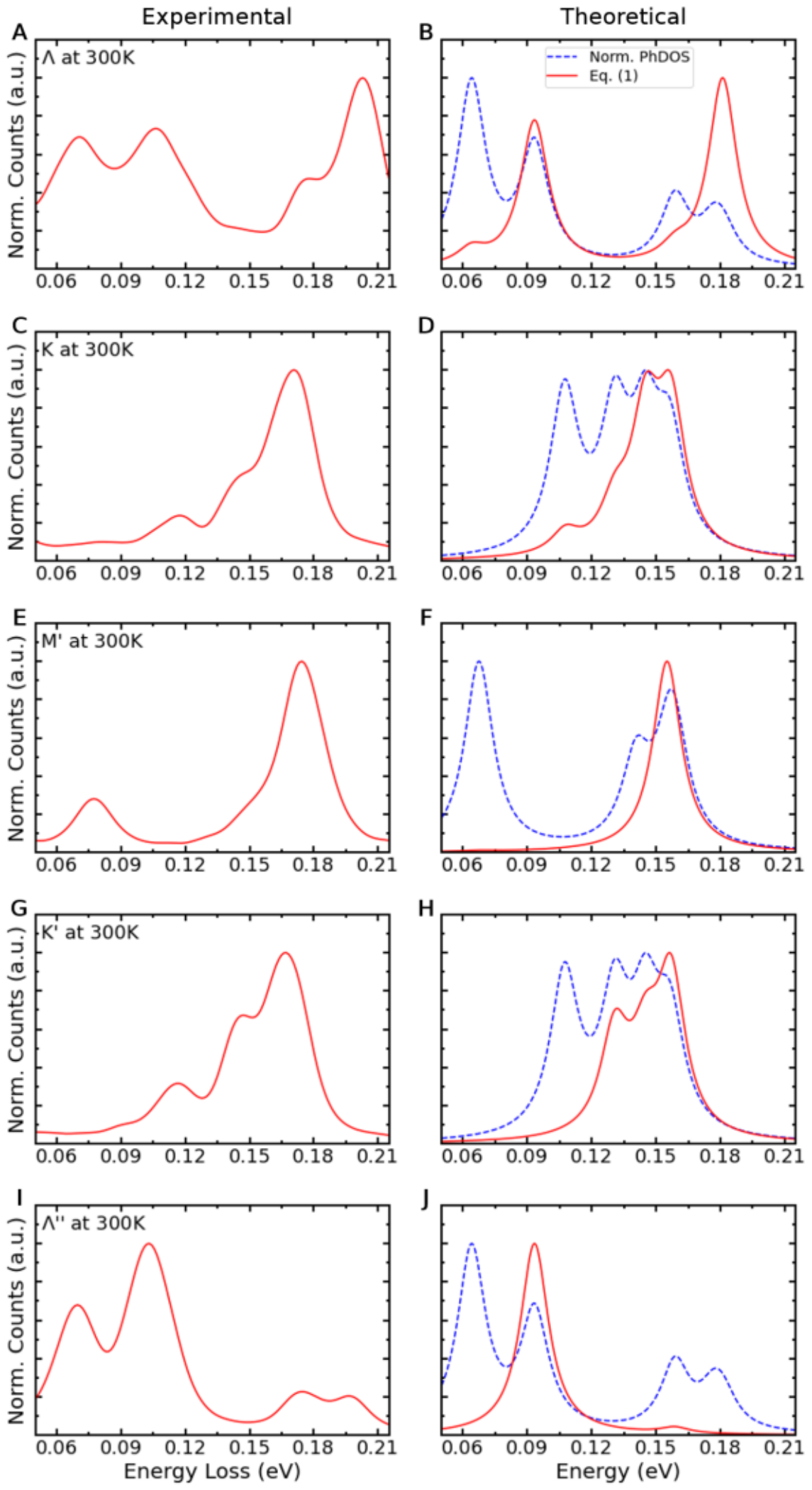

**Figure 3.** Comparison between experimental (left) and theoretical (right) EEL spectra at 300 K for momenta as labeled and discussed in the text. For the theoretical spectra, the solid lines include matrix elements as in Eq. (1), while the dashed lines are calculated with the modulus square term set to a constant, i.e., they are effective densities of phonon states at each momentum transfer. All spectra are normalized to a maximum value of unity.

There is a roughly systematic relative shift of the theoretical versus the experimental peaks, which likely arise from our choice of DFT exchange-correlation functional (we did not seek to choose a DFT functional that may give better agreement with the experimental peaks). Overall, however, the EELS relative intensities at the Λ and K symmetry points within the first BZ are reproduced well by the theory (Figures 3A,B and 3C,D), with the only exception being a low intensity in the TA-mode peak (the first peak) at the *Λ* point.

The agreement between the EELS and the theoretical relative peak intensities is not as good at momentum transfers outside the first BZ (points M′, K′, Λ″ in Figure 3, especially at the Λ″ point; see also Figure 4). Since the Stokes cross section, Eq. (1), does not include contributions from Umklapp scattering processes, we can attribute the discrepancies between theory and experiment at large momenta to the presence of strong Umklapp terms. The conclusion is corroborated by an earlier finding that Umklapp terms are necessary to simulate spatially resolved phonon spectra, which reduce momentum resolution (*35*). The Stokes theory has other limitations, e.g., it treats solids as collections of atoms that vibrate carrying a mode-independent effective charge. A complete theory should also include the effects of electron-phonon scattering on the beam itself as it traverses the sample to arrive at the detector (dynamical diffraction) (*35*–*37*) at the same level of approximation. Such an approach has recently been under development (*38*) with promising results, but such undertakings are beyond the scope of this paper.

Turning to the effect of high temperatures on the spectra at momentum transfers outside the first BZ, in Figure 4, we superpose the K/K′ and Λ/Λ′ pairs of EEL spectra at the three temperatures, 300. 800, and 1300 K. We already pointed out that the differences in each of the two pairs at 300 K (Figs. 4A and 4D demonstrate the presence of strong Umklapp processes. At the higher temperatures, the changes in the EEL spectra are quite significant. The theory of Eq. (1) is challenged even further. The calculated spectra do not change very much as a function of temperature [we confirmed that all the factors in Eq. (1) that depend on temperature change only by a small amount; see discussion in Supplementary Materials].

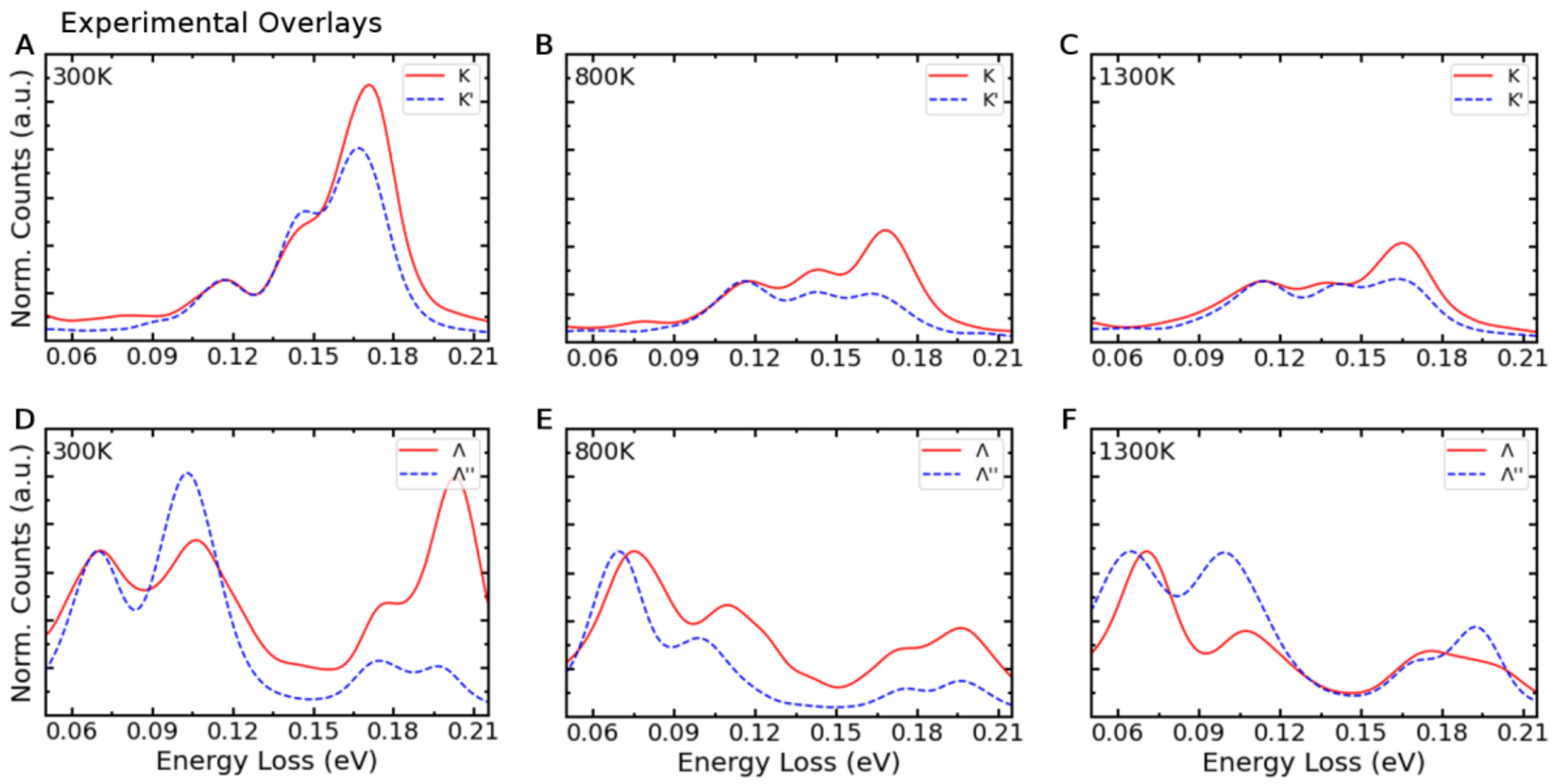

**Figure 4.** Overlay of symmetrically equivalent momenta at each temperature to illustrate differences in the experimental peak intensities. The EEL spectra at K and K′ are compared in (A)-(C) at 300 K, 800 K, and 1300 K, respectively; while the Λ and Λ′ EEL spectra are compared at 300 K, 800 K, and 1300 K in (D)-(F). The spectra in each panel are normalized at the first peak.

In summary, we have utilized EELS to measure the temperature dependence of the phonon dispersion across three BZs at very high temperatures where anharmonic effects are no longer insignificant. Extraction of the phonon modes at specific momentum transfers indicates varying degrees of phonon mode softening. The use of self-consistent phonon theory has allowed for the determination of the origin of these effects and for which modes three- and four-phonon interactions are most significant. The relative intensities of the peaks in the EEL spectra for momentum transfers within the first BZ at room temperature are well described by current theory. The EEL spectra for momentum transfers outside the first BZ differ substantially from the respective EEL spectra at equivalent points within the first BZ, which signals the presence of strong Umklapp processes. The current theory that is based on the Stokes cross section does not include Umklapp processes and has other limitations that call for a significantly more sophisticated formulation and implementation of the effect of the electron beam on the crystalline lattice and the impact of such scattering on the electron beam on its way to the detector. We anticipate that the arrival of monochromated STEM/EELS for phonons and the many successes so far with either high spatial or high momentum resolution in conjunction with theory, including the present application of the technique to high temperatures, will trigger renewed efforts to develop a more complete theory of vibrational EELS. In parallel, we anticipate that the continuing development and improvement of STEM optics, detectors, and vibration isolation will lead to improvements in the resolution of the phonon EELS spectra and improved precision in the extraction of anharmonic phonon shifts. The on-going development of cryo-STEM should also enable the ability to better understand the freeze-out of anharmonic effects at ultra-low temperatures. Such developments will improve our understanding of phonons in complex structures and their role in underpinning thermal and infrared properties of materials as well as multiphonon processes in nonradiative electronic transitions in materials.

## Acknowledgments

**Funding**: The work at Vanderbilt University was supported by the U.S. Department of Energy, Office of Science, Basic Energy Sciences, Materials Science and Engineering Directorate grant no. DE-FG02-09ER46554 and by the McMinn Endowment at Vanderbilt University. Calculations were performed at the National Energy Research Scientific Computer Center (NERSC), a U.S. Department of Energy Office of Science User Facility located at Lawrence Berkeley National Laboratory, operated under contract no. DE-AC02-05CH11231.

**Author contributions:** Conceptualization: BPS, ND, TCL, OLK. Experimental Investigation: BPS, ND, TCL, OLK. Theoretical Investigation: AO, STP. Data Analysis: AO, BPS, and JCI with contributions from all other authors. Writing – original draft: AO, STP. Writing – review & editing: AO, BPS, ND, JCI, OLK, TCL, STP.

**Competing interests:** The authors declare that they have no competing interests.

**Data and materials availability**: Reasonable requests for experimental data and theoretical results can addressed to the corresponding author.

**SUPPLEMENTARY MATERIALS**

Materials and Methods

Supplementary Text

Figures S1-S3

Supplementary Data (Tables S1-S3)

References 39-54

**Supplementary Material for**

**High-temperature phonons in h-BN: momentum-resolved vibrational spectroscopy**

Andrew O'Hara[1], Benjamin Plotkin-Swing[2], Niklas Dellby[2], Juan Carlos Idrobo[3], Ondrej L. Krivanek,[2,4] Tracy C. Lovejoy[2], and Sokrates T. Pantelides[1,5]

[1]*Department of Physics and Astronomy, Vanderbilt University, Nashville, TN 37235, USA*
[2]*Nion R&D Company, Kirkland, WA 98034, USA*
[3]*Department of Materials Science and Engineering, University of Washington, WA 98195, USA*
[4]*Department of Physics, Arizona State University, Tempe, AZ 85287, USA*
[5]*Department of Electrical and Computer Engineering, Vanderbilt University, Nashville, TN 37235, USA*

## <u>Materials and Methods:</u>

### Sample Preparation Details

Hexagonal boron nitride (h-BN) flake powder was sonicated in ethanol for declumping. A drop was placed on a Protochips Fusion Select E-chip with holey carbon film coating and allowed to evaporate. The SiC substrate of the E-chip has a 3x3 array of ~8 μm diameter holes through which the holey carbon film is visible. The E-Chip was placed in a Protochips double-tilt *in-situ* heating holder, and an ad-hoc adapter was used to make the holder compatible with the Nion U-HERMES side-entry stage.

### mc-STEM-EELS Acquisition Details

An h-BN flake of suitable thickness hanging over a hole in the film was located for the experiments. Angle-resolved electron-energy-loss spectra were acquired at three temperatures using a Nion U-HERMES with IRIS Spectrometer and DECTRIS ELA hybrid pixel detector. The energy resolution on this system at 30 kV primary energy, used for these experiments, is typically <5 meV. For these experiments, a convergence semi-angle of 2 mrad was used, with 1 pA beam current, and rectangular EELS aperture of size 0.125 x 2.0 $mm^2$. Under these conditions, the energy resolution broadened to 12-13 meV. The projector lenses were used to rotate, shift, and scale the diffraction pattern so that the aperture selected the small region of reciprocal space shown in Fig. 1E. See Ref. (*30*) for additional details on this setup. The calibration on the detector was 1.2 meV/channel in X and 0.0325 $Å^{-1}$ in Y.

There were acquisitions at three temperatures: room temperature (300 K), 800 K, and 1300 K. The sample temperature was set with the Protochips software, which uses factory calibration for the temperature of the chip. Previous experiments show close agreement between this calibration and a standardless measurement using the electron-energy gain (*17*). Each acquisition was 20 minutes in total, fractionated as 300 exposures of 4 seconds each. This resulted in a stack of 2D images, which were first aligned between frames to account for a small instrumental drift over the acquisition time, and then summed. In summed images, small wiggles due to charging of the aperture were removed by aligning the quasi-zero loss peak for each momentum point along the aperture.

### Data Analysis

Data analysis was performed using a custom Python code built on the numpy (*39*), scipy (*40*), and matplotlib (*41*) stack. After read-in, each of the raw data sets were recalibrated in order to realign the zero of energy to the primary zero-loss peak (ZLP) and to rescale the Γ-Γ′ distance to the appropriate distance. As shown in Fig. 1B-D as well as Fig. S1, this can be done via the bright spots in the lower left and right corners of the plot. When comparing the dispersion spectra in Fig. S1A-C with the theoretical dispersion (Fig. 1F), a systematic shift can be seen by inspection of the location of the TA-mode maxima at the K and K′ symmetry points and TA-mode minimum at the M′ symmetry point. Therefore, a curve-fitting extrema search was used to determine the position of the K, M′, and K′ on the previously rescaled range. A piece-wise affine transformation for each subset of the high-symmetry line was then applied to map these to the nominal values as shown Fig. S1D-E and Fig. 1B-D. At each momentum value, individual EELS line slices were then filtered using a local low rank denoising filter with a broadening of 6 meV (*42*) followed by an additional Gaussian smoothing of 4 meV. This filtering was applied for both the dispersion spectra (Fig. 1B-D) as well as all subsequent line traces (Figs. 2-4, S2). For detailed analysis of the spectra at specific momentum values, a fine-tune adjustment of the sub-dataset was done for alignment of the ZLP for that momentum. Peak deconvolution (Fig. S2) was performed using linear-least-squares fitting of multiple Lorentzian functions to an energy window subset of the data using the scipy optimize library. The energy window used for this was optimized to minimize the fit residuals while still covering all relevant peaks for a given momentum across temperatures. With Lorentzian fitting there was minimal sensitivity to minor adjustments of the range within rounding error. Since all considered modes for extraction are both significantly far from the ZPL (several multiples of the FWHM) and away from the zone center (so the ZPL is not as dominant), additional background subtraction was not performed. All fit parameters for this process are given in Table S1.

**Computational Details**

Density functional theory calculations were performed using the Vienna Ab initio Simulation Package (VASP) (*43*) with the projector-augmented wave (PAW) method (*44*, *45*) to describe core-valence interactions. For the exchange-correlation functional, the Perdew-Burke-Ernzerhof (PBE) form of the generalized gradient approximation (*46*) with an additional dispersion van-der-Waals interaction via Grimme's DFT-D3 method with Beck-Jonson damping was used (*47*, *48*). This functional has been previously shown to perform well for structural properties of h-BN (*49*). A plane-wave basis cutoff energy of 600 eV. The plane-wave basis cutoff energy was set to 600 eV and the Brillouin zone was sampled with a $10 \times 10 \times 4$ Γ-centered k-point grid for the primitive cell. The optimized lattice constants at 0 K were determined to be $a = 2.5066$ Å, $c = 6.5605$ Å. Thermal lattice expansion was included for each temperature by using available experimental (*23*, *24*) and theoretical thermal expansion coefficients (*25*). The resulting lattice constants used for subsequent calculations were: $a = 2.5055$ Å, $c = 6.6104$ Å at 300 K; $a = 2.5036$ Å, $c = 6.7414$ Å at 800 K; and $a = 2.5033$ Å, $c = 6.8820$ Å at 1300 K. At each temperature, density-functional perturbation theory was used to determine the dielectric tensor and Born effective charges needed for inclusion of the non-analytic contribution to the phonon dispersion. Phonon calculations were performed using the ALAMODE package (*50*). Quasi-harmonic (2$^{nd}$ order) force constants were extracted by performing least squares fits to 0.01 Å displacements in a $5 \times 5 \times 2$ supercell of the primitive unit cells. Cubic and quartic anharmonic force constants were extracted using the compressive sensing approach (*51*). First, ab-initio molecular dynamics calculations with lower accuracy parameters (500 eV cutoff and Γ-point only Brillouin zone sampling) were performed with a time-step of 0.5 femtoseconds at 500 K for 2,500 steps. Fifty evenly spaced structures were taken from the run and an additional randomly-oriented, 0.1 Å in magnitude displacement was added to each atom. For each generated structure, a DFT calculation at the original accuracy was performed to calculate the interatomic forces. The anharmonic force constants were then determined using the least absolute shrinkage and selection operator (LASSO) technique with four-fold cross validation. A nearest-neighbor cutoff radius of 9 Bohr (4.76 Å) was used to limit the total number of force constants. Such a cutoff includes up to 5$^{th}$ nearest-neighbors in-plane and 4$^{th}$ nearest-neighbors in the adjacent plane. For both the cubic and quartic terms, both three-phonon and four-phonon interactions were included. The temperature-dependent renormalization of the phonon band structure was performed within the framework of the self-consistent phonon theory (SCP) which includes cubic tadpole terms and quartic loop terms in the many-body diagrammatic expansion of the phonon self-energy (*31*). A primary k-point grid commensurate with the supercell ($5 \times 5 \times 2$) was used and the self-consistent equations solved on a denser interpolation grid ($15 \times 15 \times 6$). Additional temperature-dependent cubic renormalization of the phonon band structure via the cubic bubble diagrammatic term was included by solving the nonlinear quasiparticle (QP-NL) equation for the phonon self energy on top of the first-order SCP solution (*32*). Additional properties needed for computing the EELS cross-section as in main text Eq (1), such as the eigenvectors or mean-square displacement (msd), are calculated using the renormalized dynamical matrix. The msd values for 300K and 1300K used in the main text are summarized in Table S3. To obtain effective charges, we used the effective charges from Ref. (*15*) extracted via the Engauge Digitizer program.

## Supplementary Text

### EELS as a probe of sample temperature

One other feature that can be obtained from EEL phonon spectra such as those of Figures 1B-D and 2 is the temperature of the material, as demonstrated in Ref. (*17*). According to the principle of detailed balance (*52*), the ratio of the transition probabilities from a low- to a high-energy vibrational state and its reverse, corresponding to the ratio of the energy-gain and energy-loss EELS peak intensities, is equal to the Boltzmann factor:

$$I_{gain}/I_{loss} = e^{-E_{ph}/k_B T}, \tag{S1}$$

where $E_{ph}$ is the energy of the pertinent phonon. In Ref. (*17*), it was shown that the STEM can function as a thermometer by measuring the ratio $I_{gain}/I_{loss}$ for a single phonon energy over a wide range of temperatures and showing that the temperatures extracted from Eq. (S1) are in excellent agreement with the corresponding nominal temperatures. Here, we used the EEL spectra like those of Figure 2 for all the phonon energies in the dispersions of Figures 1B-D at the three considered temperatures. Figures S2A-C show heat maps of $T - T_{nom}$ , where $T$ is the temperature extracted from Eq. (S1) at each mode in the dispersion and $T_{nom}$ is the nominal temperature (300, 800 or 1300 K). As per the scale bar, the color white indicates $T - T_{nom} = 0$, *i.e.*, excellent agreement between the temperature extracted from virtually all phonon modes and the corresponding nominal temperature of the sample. The agreement is good even for the 300-K measurements in which the energy-gain part of the spectrum is very weak. Slight deviations ($T < T_{nom}$, orange color) are visible in the TA phonons at the M′ point and near the Γ points at 800 K. Somewhat larger deviations at both the TA and LA phonon modes are visible at 1300 K, where higher levels of noise are evident. The deviations arise because, at these temperatures, the thermal energy $k_B T$ (67 and 108 meV at 800 and 1300 K, respectively) is roughly at or at a higher level than the pertinent phonon energies, whereby noise and other loss processes may contribute to the EEL spectra. We conclude that phonon EELS can be used to extract temperatures at the nanoscale as long as the phonon modes that are used are sufficiently away from the Γ points and have energies above the expected nominal temperature.

**Temperature dependence of the Stokes cross section** [Equation (1) in the main text]

There are several terms that exhibit a temperature dependence: the Bose-Einstein occupation, the phonon mode energies, the Debye-Waller factors, the eigenvectors, and the effective charges.

Using the four phonon energies that correspond to the four peaks in the spectra, the ratios of the Bose-Einstein occupation factors $(1 + n_{q\nu})$ at the two temperatures are 1.6, 1.5, 1.4, and 1.3, respectively. While these ratios imply an enhancement of the intensities of the two acoustic modes relative to those of the two optical modes, the predicted relative enhancement is not nearly as large as those in the EEL spectra. Using the same modes, the ratios of the ${\omega_{\boldsymbol{q}\nu}}^{-1}$ terms at different temperatures indicate enhancements of order 1.02, 1.01, 1.02, and 1.02, respectively. The inconsequential nature of this enhancement is due to the order of magnitude of the phonon energy shifts compared to the phonon energies themselves. Restricting to the in-plane atomic motions, we considered a Debye-Waller factor (DWF) of the form $\exp(-q^2\langle u^2\rangle/2)$ where the $\langle u^2\rangle$ are the calculated in-plane mean-square displacements (see Table S3 for values). Although the mean-square displacements increase for both boron and nitrogen, they do so in tandem and are relatively similar, indicating an increase in the DWF from 1.02 to 1.05 for both boron and nitrogen. when increasing the temperature from 300 K to 1300 K at the $K'$ point. Although there is a minor variation in the eigenvectors across temperatures, no significant changes in the plotted spectra occur when changing only the eigenvectors either.

The final term to consider for temperature-induced changes and the only term with a significant $q$ dependence that would be different for different atomic species is the effective charge whether it be based on an atomic form factor (*14*) or dynamical charge (*15*). In the theory plots of Fig. 3, we utilized the same charges as in Ref. (*15*) where the long wavelength limit ($q \rightarrow \infty$) is the ionic charge and the short wavelength limit ($q \rightarrow 0$) is the screened Born effective charge. As the temperature increases, the long wavelength limit cannot change. In the short wavelength limit, while the larger atomic displacements may change the dynamic Born effective charges slightly from the limit of linear response, the change in the screening would be more significant. At high temperatures, narrowing of the bandgap and an increase in the number of thermal electron-hole pairs would, generally, lead to an increase in the dielectric constant which has been reported experimentally for h-BN (*53*). Hence a decrease of the effective charges from the present values ($\sim \pm 0.5$) towards zero would occur and the overall functional form of the effective charge as plotted in Figs. 2b and Extended Data Fig. 6 of Ref. (*15*) would likely change very little. Therefore, at the current level of theory, the effective charges also cannot properly describe the experimentally observed temperature dependence. One possible improvement would be to include explicit electron-phonon coupling, which is often phonon mode dependent, especially in polar materials (*54*).

## Supplementary Figures

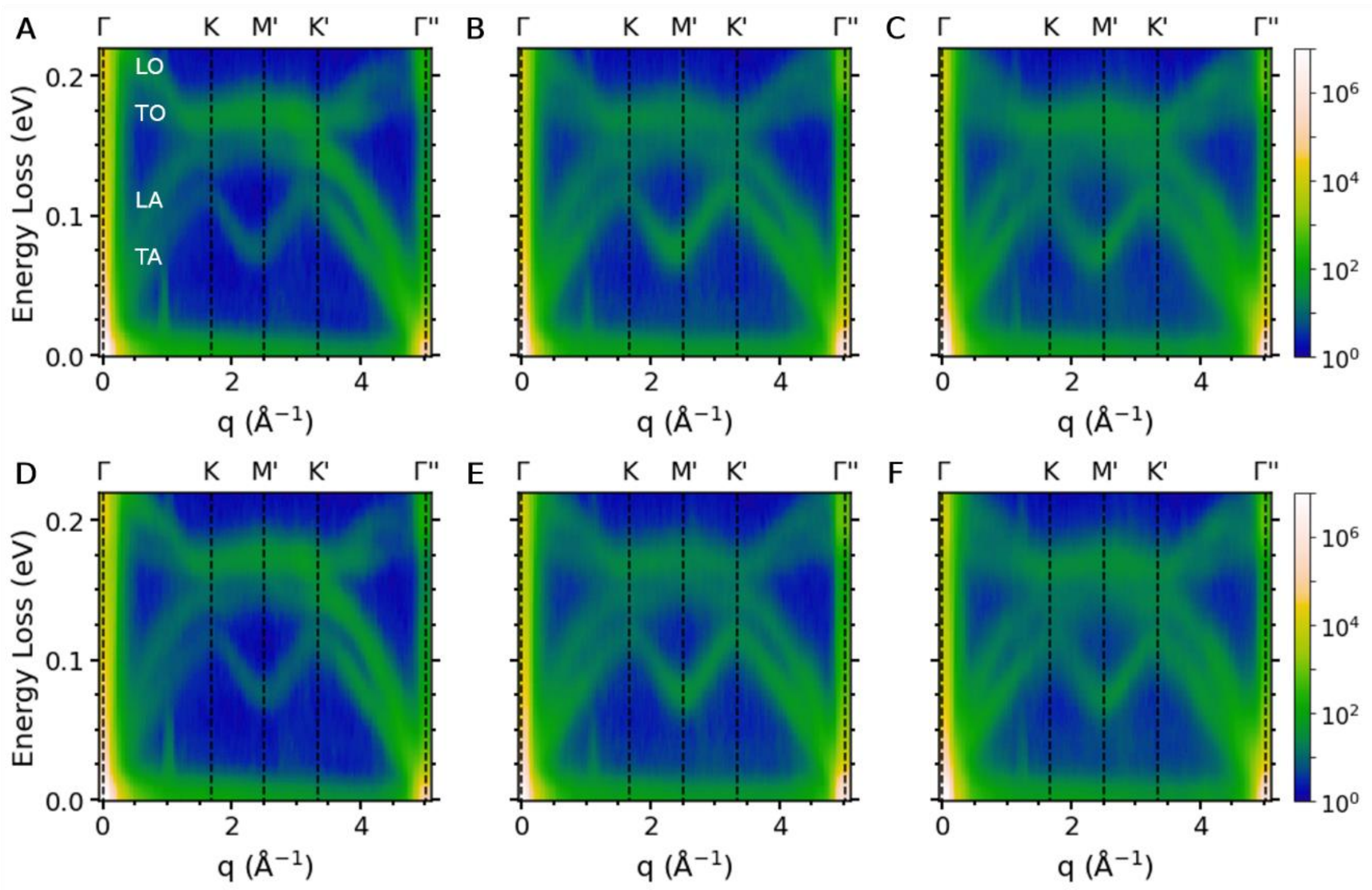

**Figure S1.** STEM-EELS derived phonon dispersions at 300 K (A,D), 800 K (B,E), and 1300 K (C,F) showing the as-collected and initially scaled data (A-C) along with the affine transformed momentum axis used for Figure 2 (D-F). Details of this scaling and transformation are described in the methods section.

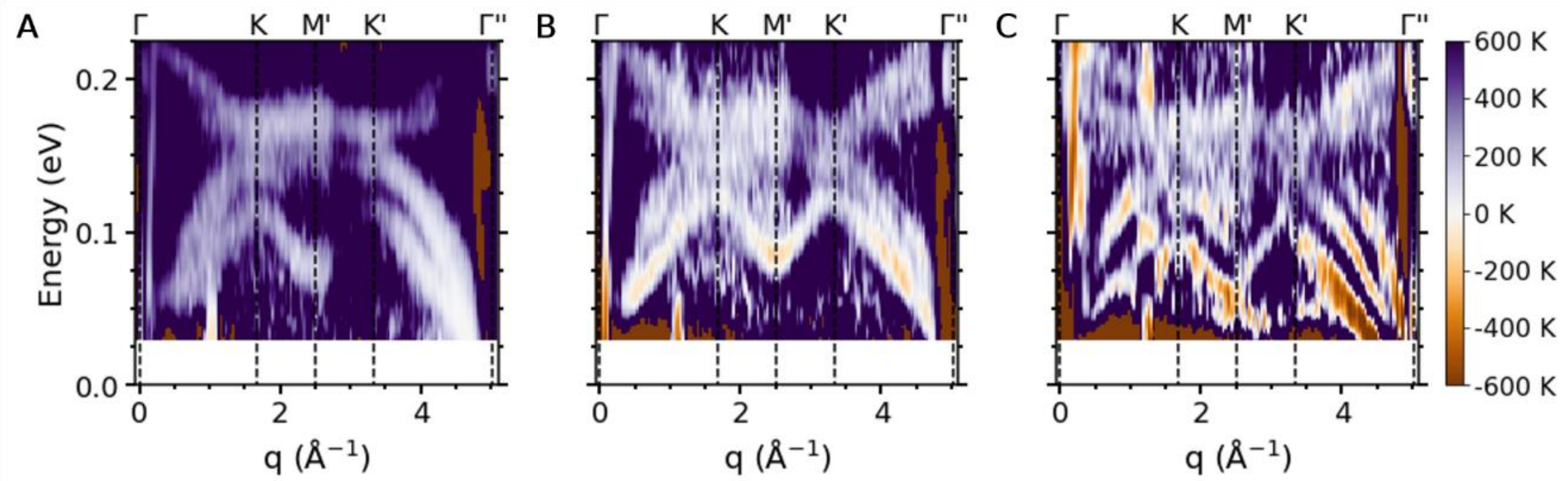

**Figure S2.** Temperature extracted from the energy loss and gain spectra (shown in Figs. 2A-C and 3) relative to the nominal measured temperatures at (**A**) 300, (**B**) 800, and (**C**) 1300 K, respectively.

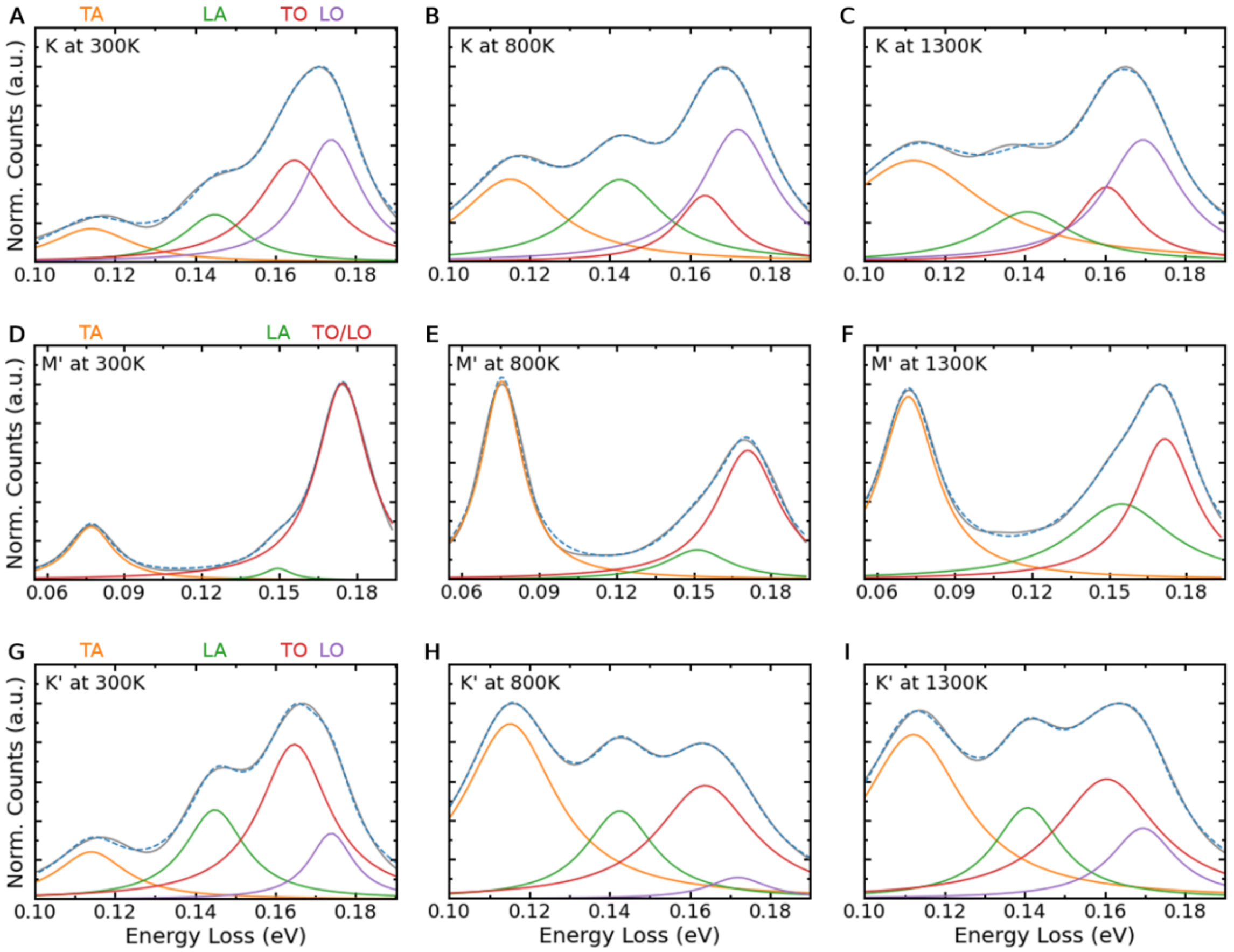

**Figure S3.** Lorentzian peak deconvolution for the K point at 300 K (**A**), 800 K (**B**), and (**C**) 1300 K. Peak deconvolution for the M′ point at 300 K **(D)**, 800 K **(E)**, and 1300 K **(F)**. Peak deconvolution for the K′ at 300 K **(G)**, 800 K **(H**), and 1300 K (**I**). For the K and K′ points, the peak-energy deconvolution was constrained to obey symmetry. Note that the normalization of Fig. 3 is not retained in these plots.

**Supplementary Data**

**Table S1.** Phonon energies (in meV) extracted from EEL spectra at high-symmetry BZ points at three temperatures. Due to the closeness in energies, only a combined TO/LO value is given for the M′ point. Details of the fitting procedure to extract the phonon energies is given in the Data Analysis section of the Supplementary Materials.

| Momentum | Temperature | TA | LA | TO | LO |
|---|---|---|---|---|---|
| K/K′ | 300 K | 114 | 145 | 165 | 174 |
| | 800 K | 115 | 142 | 164 | 172 |
| | 1300 K | 112 | 141 | 160 | 170 |
| M′ | 300 K | 77 | 149 | 174 | |
| | 800 K | 75 | 151 | 171 | |
| | 1300 K | 72 | 155 | 171 | |

**Table S2.** All fit parameters for the deconvolution of the peaks in Figure 3 of the main text. All energies (energy range, mode energy, and mode $\gamma$) are in eV.

| Momentum | Temp. | Range Min | Range Max | Phonon Mode | Mode Energy | Mode Amplitude | Mode γ | Fit Residual |
|---|---|---|---|---|---|---|---|---|
| K | 300 K | 0.100 | 0.190 | TA | 0.114 | 0.007 | 0.012 | 0.018 |
| | | | | LA | 0.145 | 0.007 | 0.010 | |
| | | | | TO | 0.165 | 0.018 | 0.011 | |
| | | | | LO | 0.174 | 0.018 | 0.009 | |
| | 800 K | 0.100 | 0.190 | TA | 0.115 | 0.020 | 0.015 | 0.002 |
| | | | | LA | 0.142 | 0.017 | 0.013 | |
| | | | | TO | 0.164 | 0.009 | 0.008 | |
| | | | | LO | 0.172 | 0.024 | 0.011 | |
| | 1300 K | 0.100 | 0.190 | TA | 0.112 | 0.035 | 0.022 | 0.008 |
| | | | | LA | 0.141 | 0.011 | 0.014 | |
| | | | | TO | 0.160 | 0.011 | 0.009 | |
| | | | | LO | 0.170 | 0.022 | 0.012 | |
| M′ | 300 K | 0.055 | 0.195 | TA | 0.077 | 0.009 | 0.010 | 0.011 |
| | | | | LA | 0.149 | 0.001 | 0.006 | |
| | | | | TO/LO | 0.174 | 0.039 | 0.012 | |
| | 800 K | 0.055 | 0.198 | TA | 0.075 | 0.031 | 0.010 | 0.040 |
| | | | | LA | 0.151 | 0.007 | 0.015 | |
| | | | | TO/LO | 0.171 | 0.031 | 0.015 | |
| | 1300 K | 0.055 | 0.198 | TA | 0.072 | 0.039 | 0.013 | 0.014 |
| | | | | LA | 0.155 | 0.029 | 0.024 | |
| | | | | TO/LO | 0.171 | 0.032 | 0.014 | |
| K′ | 300 K | 0.100 | 0.190 | TA | 0.114 | 0.008 | 0.010 | 0.011 |
| | | | | LA | 0.145 | 0.013 | 0.009 | |
| | | | | TO | 0.165 | 0.027 | 0.011 | |
| | | | | LO | 0.174 | 0.007 | 0.007 | |
| | 800 K | 0.100 | 0.190 | TA | 0.115 | 0.039 | 0.014 | 0.002 |
| | | | | LA | 0.142 | 0.015 | 0.010 | |
| | | | | TO | 0.164 | 0.027 | 0.015 | |
| | | | | LO | 0.172 | 0.003 | 0.009 | |
| | 1300 K | 0.100 | 0.190 | TA | 0.112 | 0.041 | 0.015 | 0.007 |
| | | | | LA | 0.141 | 0.015 | 0.010 | |
| | | | | TO | 0.160 | 0.030 | 0.015 | |
| | | | | LO | 0.170 | 0.011 | 0.010 | |

**Table S3.** Mean-square-displacement values in Å$^2$ used for the calculation of the Debye-Waller factors in Eq. (1) of the main text.

| Temperature | Boron | | Nitrogen | |
|---|---|---|---|---|
| | $\langle\mu_x^2\rangle = \langle\mu_y^2\rangle$ | $\langle\mu_z^2\rangle$ | $\langle\mu_x^2\rangle = \langle\mu_y^2\rangle$ | $\langle\mu_z^2\rangle$ |
| 300 K | $2.98 \times 10^{-3}$ | $1.76 \times 10^{-2}$ | $2.71 \times 10^{-3}$ | $2.05 \times 10^{-2}$ |
| 1300 K | $8.97 \times 10^{-3}$ | $6.51 \times 10^{-2}$ | $8.50 \times 10^{-3}$ | $7.78 \times 10^{-2}$ |